\documentclass{article}

\usepackage[english]{babel}

\usepackage[letterpaper,top=2cm,bottom=2cm,left=3cm,right=3cm,marginparwidth=1.75cm]{geometry}

\usepackage{amsmath}
\usepackage{graphicx}
\usepackage[colorlinks=true, allcolors=blue]{hyperref}
\usepackage{booktabs}

\title{Commentary to Wan et al. (2014):\\
Estimating the standard deviation \\
from the sample size and range or quartiles}
\author{Massimo Borelli \footnote{\texttt{massimo.borelli@unicz.it}, UMG School of PhD Programmes, University \textit{Magna Gr\ae cia} of Catanzaro.} , ORCID: 0000-0002-6461-6563}

\begin{document}

\maketitle

$$ $$
Mathematics Subject Classification (MSC): 62P10 

$$ $$

\begin{abstract}
This short note proposes two additive corrections to a pair of relations published in Wan et al. \cite{wan2014estimating} in order to extend them to a 'small sample size' condition. In particular we focus the interest on the possibility to provide an estimate to the sample standard deviation $\sigma$ when knowing only the sample size $n$, the range $[a, b]$ and/or the quartiles $Q_1, Q_3$ of some data. Our results allow to explicitely compute $\sigma$, for instance with software \textsf{R} or any spreadsheet, for any sample size $n \geq 2$. \textsf{R} codes and data are publicly available on \url{https://github.com/MassimoBorelli/sd}

\end{abstract}

$$ $$

\section{Background}

In their 2014 \textit{BMC medical research methodology} paper \cite{wan2014estimating}, Xiang Wan and colleagues improve a previous work by Stela Pudar Hozo et al. \cite{hozo2005estimating}, appeared on the same journal. The focus of both works concern the possibility to estimate the sample mean and the sample standard deviation knowing only the sample size, the median, the range and/or the interquartile range of the data. Such kind of arguments assumes particular relevance in experimental design, in systematic reviewing or during meta-analysis investigations. As an example, suppose one is interested to establish a proper sample size when designing a prospective study in which repeated measures anova will be addressed: a typical relation (Chow et al. \cite{chow2017sample}, Chapter 15; here $\Delta $ represents difference between means) to use could be:

$$ n \geq \frac{2 \sigma^2 (z_{\alpha/2} + z_{\beta})^2}{\Delta^2} $$

\noindent Unfortunately, when literature results are reported in a non-parametric way, it is tricky to guess means $\mu$ and standard deviations $\sigma$ in order to apply such kind of formulas. In their paper, Wan et al. face up three typical scenarios of not-parametric descriptive statistics reported:


\begin{itemize}
  \item[$\mathbf{C}_1$] the median $m$, the range $\left[ a, b \right]$ and the sample size $n$
  \item[$\mathbf{C}_2$] the median $m$, the range $\left[ a, b \right]$, the quartiles $\left[ Q_1, Q_3 \right]$ and the sample size $n$
  \item[$\mathbf{C}_3$] the median $m$, the quartiles $\left[ Q_1, Q_3 \right]$ and the sample size $n$
\end{itemize}

\noindent In the following subsection we summarize their results.

\subsection{Estimating means.}

\noindent According to the recalled scenarios,  Wan and colleagues \cite{wan2014estimating} publish three relationships devoted to estimate the sample mean. Some formulas have originally been obtained by Hozo et al. \cite{hozo2005estimating} (scenario [$\mathbf{C}_1$]) and by Martin Bland in \cite{bland2015estimating} (scenario [$\mathbf{C}_2$]), exploiting straightforward algebraic considerations and basic inequalities. Here we report their findings concerning $\mu$'s:

\begin{itemize}
  \item[$\mathbf{C}_1$] $\mu \approx  \frac{a + 2m + b}{4} + \frac{a - 2m + b}{4n} \approx  \frac{a + 2m + b}{4} $ 
  \item[$\mathbf{C}_2$] $\mu \approx  \frac{a + 2Q_1 + 2m + 2Q_3 + b}{8} $
  \item[$\mathbf{C}_3$] $\mu \approx  \frac{Q_1 + m + Q_3}{3} $
\end{itemize}

\noindent In their paper, authors also propose some simulations enlighting the possible error occurring in estimating the sample mean in several (artificial, random generated) normal and non-normal data.

\subsection{Estimating standard deviations.}

The novelty in Wan et al. \cite{wan2014estimating} concerns the estimation of standard deviations $\sigma$, according to the following statements:

$$\sigma \approx  \frac{b - a}{\xi(n)}  $$

$$\sigma \approx  \frac{Q_3 - Q_1}{\eta(n)}  $$

$$\sigma \approx  \frac{1}{2} \left(     \frac{b - a}{\xi(n)} +   \frac{Q_3 - Q_1}{\eta(n)}   \right)  $$

\noindent respectively on scenario $\mathbf{C}_1$, $\mathbf{C}_3$, and $\mathbf{C}_2$. Despite such an elegant and simple appearance, the computations which lead authors to obtain the two novel real valued functions $\xi(n)$ and $\eta(n)$  are not straightforward at all. The difficulties are related  to the non-integrability of both the probability density function $\phi(z) = \frac{1}{\sqrt{2 \pi}} \exp(-z^2/2) $ and the cumulative distribution function $\Phi(z) = \int_{-\infty}^{z}\phi(t)dt $ of the standard normal distribution, which are involved in evaluating the expected values of certain appropriate order statistics.To overcome the problem of non-integrability, the authors distinguish two cases.

\subsubsection{Estimating standard deviations with 'small' sample sizes.}
\label{nlth50}

In case of sample sizes between 1 and 50, Wan et al. resorted the numerical integrator routines implemented in \textsf{R} \cite{R} and reported the numerical evidences in the following Tables \ref{tableXi} and \ref{tableEta}. We stress here the focus point: the values hereby listed are only numerically computed values, but not any explicit formula yielding such results is known.

\begin{table}[ht]
	\centering
	\caption{Numerical values of  $\xi(n)$  when $n \leq 50$, according to  Wan et al. }
	\label{tableXi}
	{
		\begin{tabular}{|cc|cc|cc|cc|cc|}
			\toprule
			n & $\xi(n)$ & n & $\xi(n)$ & n & $\xi(n)$ & n & $\xi(n)$ & n & $\xi(n)$   \\
			\cmidrule[0.4pt]{1-10}
1 & 0 & 11 & 3,173 & 21 & 3,778 & 31 & 4,113 & 41 & 4,341 \\
2 & 1,128 & 12 & 3,259 & 22 & 3,819 & 32 & 4,139 & 42 & 4,361 \\
3 & 1,693 & 13 & 3,336 & 23 & 3,858 & 33 & 4,165 & 43 & 4,379 \\
4 & 2,059 & 14 & 3,407 & 24 & 3,895 & 34 & 4,189 & 44 & 4,398 \\
5 & 2,326 & 15 & 3,472 & 25 & 3,931 & 35 & 4,213 & 45 & 4,415 \\
6 & 2,534 & 16 & 3,532 & 26 & 3,964 & 36 & 4,236 & 46 & 4,433 \\
7 & 2,704 & 17 & 3,588 & 27 & 3,997 & 37 & 4,259 & 47 & 4,450 \\
8 & 2,847 & 18 & 3,640 & 28 & 4,027 & 38 & 4,280 & 48 & 4,466 \\
9 & 2,970 & 19 & 3,689 & 29 & 4,057 & 39 & 4,301 & 49 & 4,482 \\
10 & 3,078 & 20 & 3,735 & 30 & 4,086 & 40 & 4,322 & 50 & 4,498 \\
			\bottomrule
		\end{tabular}
	}
\end{table}

\begin{table}[ht]
	\centering
	\caption{Numerical values of  $\eta(n)$  when $n \leq 50$, according to  Wan et al. }
	\label{tableEta}
	{
		\begin{tabular}{|cc|cc|cc|cc|cc|}
			\toprule
			n & $\eta(n)$ & n & $\eta(n)$ & n & $\eta(n)$ & n & $\eta(n)$ & n & $\eta(n)$   \\
			\cmidrule[0.4pt]{1-10}
1 & 0,990 & 11 & 1,307 & 21 & 1,327 & 31 & 1,334 & 41 & 1,338 \\
2 & 1,144 & 12 & 1,311 & 22 & 1,328 & 32 & 1,334 & 42 & 1,338 \\
3 & 1,206 & 13 & 1,313 & 23 & 1,329 & 33 & 1,335 & 43 & 1,338 \\
4 & 1,239 & 14 & 1,316 & 24 & 1,330 & 34 & 1,335 & 44 & 1,338 \\
5 & 1,260 & 15 & 1,318 & 25 & 1,330 & 35 & 1,336 & 45 & 1,339 \\
6 & 1,274 & 16 & 1,320 & 26 & 1,331 & 36 & 1,336 & 46 & 1,339 \\
7 & 1,284 & 17 & 1,322 & 27 & 1,332 & 37 & 1,336 & 47 & 1,339 \\
8 & 1,292 & 18 & 1,323 & 28 & 1,332 & 38 & 1,337 & 48 & 1,339 \\
9 & 1,298 & 19 & 1,324 & 29 & 1,333 & 39 & 1,337 & 49 & 1,339 \\
10 & 1,303 & 20 & 1,326 & 30 & 1,333 & 40 & 1,337 & 50 & 1,340 \\
			\bottomrule
		\end{tabular}
	}
\end{table}

\subsubsection{Estimating standard deviations with 'large' sample sizes.}

When sample sizes are higher than 50, in order to approximate $\xi(n)$ and $\eta(n)$  Wan et al. employ a method  described in Gunnar Blom \cite{blom1958statistical}: such a method require to compute the upper z-th standard normal percentile $\Phi^{-1}(p)$ function, which for instance in \textsf{R} is commonly retrieved by the command \texttt{qnorm()}:

\begin{equation}
\label{xiphi}
\xi(n) \approx {2 \cdot \Phi^{-1} \left( \frac{n-0.375}{n+0.25} \right) } 
\end{equation}

\begin{equation}
\label{etaphi}
\eta(n) \approx {2 \cdot \Phi^{-1} \left( \frac{0.75 n - 0.125}{n+0.25}  \right) } 
\end{equation}

\subsection{Our research question}

Our research question concerns the possibility to adapt the asymptotic relations (\ref{xiphi}) and (\ref{etaphi}) in order to extend them also in the 'small' sample sizes case $n \leq 50$ described in subsection \ref{nlth50}. In Section \ref{Results}, two novel functions $\delta(n)$ and $\varepsilon(n)$ are introduced and statistically estimated by $\hat \delta(n)$ and $\hat \xi(n)$, in order to provide two explicit functions which mimic $\xi(n)$ and $\eta(n)$ behaviour also when $n \leq 50$:

\begin{equation}
\label{XiPhiDelta}
\hat \xi(n) \equiv {2 \cdot \Phi^{-1} \left( \frac{n-0.375}{n+0.25} \right) + \hat \delta(n)} 
\end{equation}

\begin{equation}
\label{EtaPhiVarepsilon}
\hat \eta(n) \equiv {2 \cdot \Phi^{-1} \left( \frac{0.75 n - 0.125}{n+0.25}  \right) + \hat \varepsilon(n)} 
\end{equation}

\noindent With our addictive corrections, for instance, it is possible to improve the results provided in the Excel spreadsheet (Additional file 2) published by Wan et al. in their supplementary materials. In fact, their spreadsheet disregard the two cases $2 \leq n \leq 50$ and  $n > 50$, exploiting the same estimations (\ref{xiphi}) and (\ref{etaphi}) for both kind of sample dimension $n$, 'small' or 'large'.

\newpage
\section{Results}
\label{Results}

\begin{figure}[ht] 
\label{Arxiv1Plot1}
	\centering
		\includegraphics[scale = 0.4]{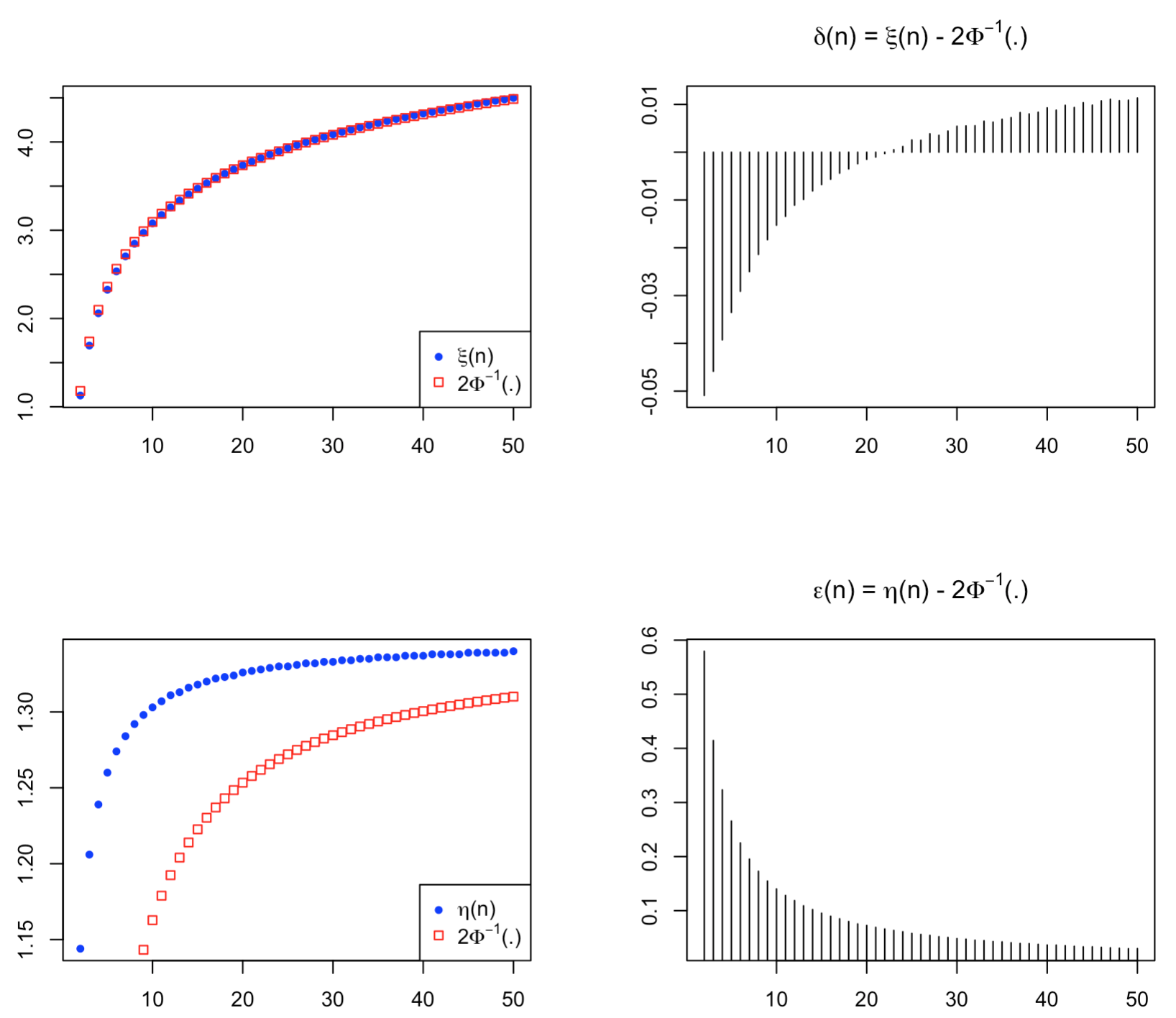} 
\caption{\small in the upper left panel, the blue bullets represent the $\xi(n)$ values as numerically computed in Table \ref{tableXi}, while the red squares depict the function ${2 \cdot \Phi^{-1} \left( \frac{n-0.375}{n+0.25} \right) }$.  On the upper right panel, the residuals $\delta(n)$ are plotted. In the lower left panel,  the blue bullets represent function $\eta(n)$ as listed in Table \ref{tableEta}, while $2 \cdot \Phi^{-1} \left( \frac{0.75 n - 0.125}{n+0.25}  \right) $ are plotted with red squares. In the lower right panel, sticks describe residuals $\varepsilon(n)$.}
\end{figure}

In what follows, we assume that $n \geq 2$, being the case $n = 1$ not practically relevant. All the following analyses are detailed in  \url{https://github.com/MassimoBorelli/sd}. To start, one defines:

$$ \delta(n) := \xi(n) - 2 \cdot \Phi^{-1} \left( \frac{n-0.375}{n+0.25} \right)   $$
$$ \varepsilon(n) := \eta(n) - 2 \cdot \Phi^{-1} \left( \frac{0.75 n - 0.125}{n+0.25}  \right) $$

\noindent where $\xi(n)$ and $\eta(n)$ are the tabulated values in Tables \ref{tableXi} and \ref{tableEta} and $\Phi^{-1}$ is the upper z-th standard normal quantile function. A simple visual inspection to the above right sides plots shows that $\varepsilon(n)$ and $\delta(n)$ approximately differs by one order of magnitude. This is the reason why we start discussing $\varepsilon(n)$ within scenario $\mathbf{C}_3$.

\subsection{Construction of epsilon(n)}

Let us observe that $\varepsilon(n) > 0, \forall n \in \mathbf{N}$; therefore it will be possible to consider logarithms. We claim that it is worth seeking two constants $a, b < 0$ in order to set

$$ \hat \varepsilon(n; a, b) := \exp \left(   \frac{n}{a+b \cdot n}   \right) > 0$$

\noindent as a possible estimate of $\varepsilon(n)$. In fact, if the above statement hold, we would have that:

$$ 0 > \frac{n}{a+b \cdot n}  = \log \left(  \hat  \varepsilon(n) \right) $$

$$  \frac{n}{\log \left(  \hat  \varepsilon(n) \right) }  = a+b \cdot n < 0$$

\noindent and therefore the change of variable $Y = {n}/{\log \left(  \hat  \varepsilon(n) \right) } $ would yield to a linear relation:

$$ Y  = a+b \cdot n $$

\bigskip

\begin{figure}[ht] 
\label{Arxiv1Plot2}
	\centering
		\includegraphics[scale = 0.4]{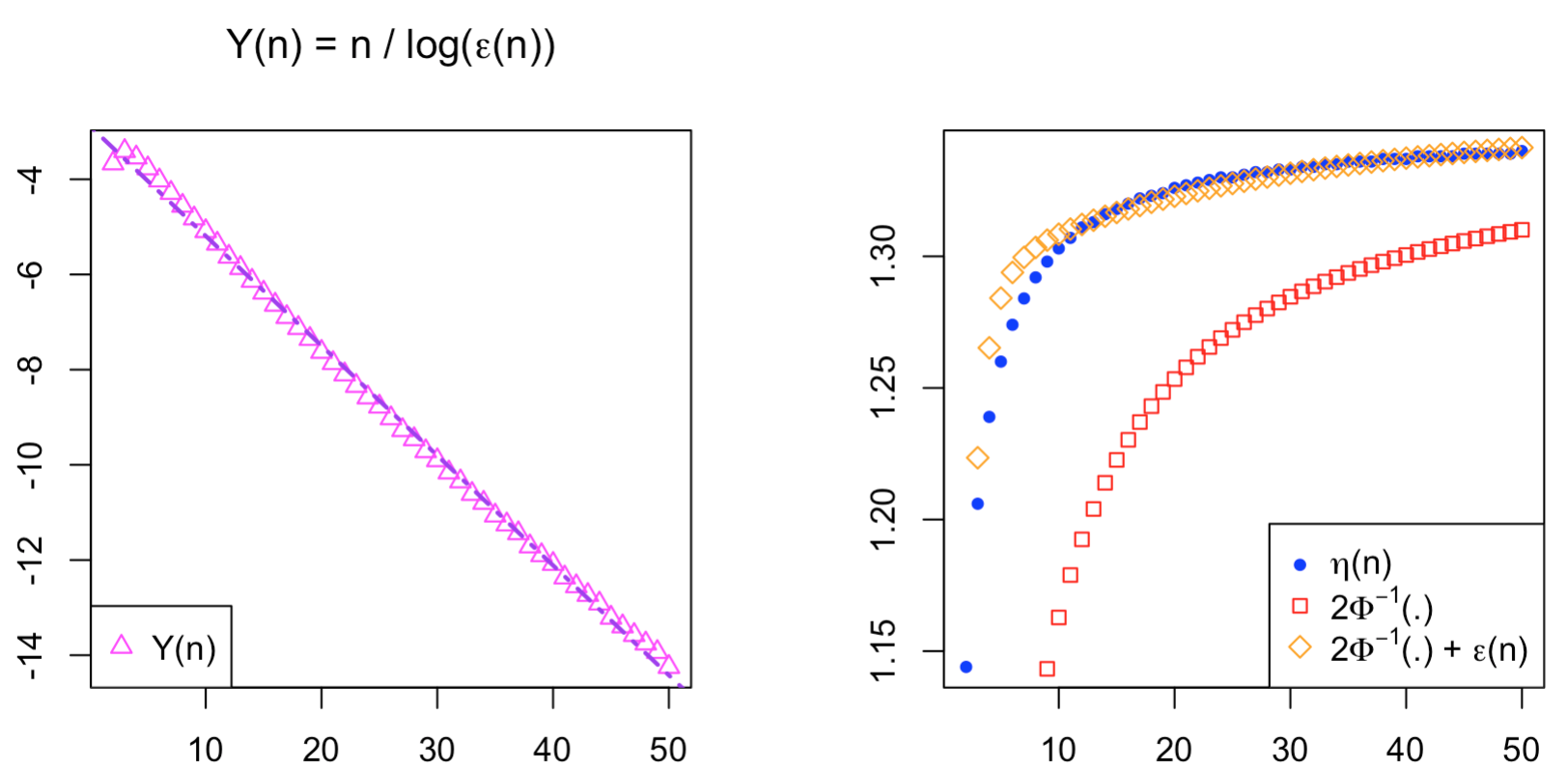} 
\caption{ on the left, the linear behaviour of $Y(n)$ in magenta triangles; the dashed purple regression line is estimated. On the left, the blue bullets represent the $\xi(n)$ values, the red squares depict the function Wan et al., ${2 \cdot \Phi^{-1} \left( \frac{0.75 n - 0.125}{n+0.25} \right) }$, while the orange diamonds plot our estimated proposal ${2 \cdot \Phi^{-1} \left( \frac{0.75 n - 0.125}{n+0.25} \right) } + \exp \left( \frac{n}{-2.882 - 0.231 \cdot n} \right)$. }
\end{figure}

As shown in Figure \ref{Arxiv1Plot2}, if we plot $n / \log(\varepsilon(n))$ versus $n$, the pink triangles  enhance the linear behaviour of $Y(n)$, justifying our initial claim: with a residual standard error of 0.141 and a determination coefficient $R^2 = 0.998$ the points appears to lie on the  least mean squares $Y = -2.882 - 0.231 \cdot n$  regression line. The summary below reported is an adaptation of the one provided by the \texttt{lm} function of \textsf{R} \cite{R}:

\begin{table}[ht]
\centering
\begin{tabular}{rrrrr}
  \hline
 & Estimate & Std. Error & t value & Pr($>$$|$t$|$) \\ 
  \hline
 $a $ & -2.8822 & 0.0421 & -68.48 & $<$ 0.001 \\ 
  $b$ & -0.2308 & 0.0014 & -162.30 & $<$ 0.001 \\ 
   \hline
\end{tabular}
\end{table}

Consequently, substituting $a$ and $b$ according to previous passages, we obtain a plausible estimate for $\varepsilon(n)$:

\begin{equation}
\label{ourVarepsilon}
 \hat \varepsilon(n) := \exp  \left( \frac{n}{-2.882 - 0.231 \cdot n} \right)
\end{equation}

\noindent and therefore the $\eta(n)$ can be better approximated, according to the following relation:

$$ \hat \eta (n) = {2 \cdot \Phi^{-1} \left( \frac{0.75 n - 0.125}{n+0.25} \right)  + \exp \left( \frac{n}{-2.882 - 0.231 \cdot n} \right) }$$

\noindent leading to the conclusive relation to estimate the standard deviation $\sigma$ within scenario $\mathbf{C}_3$:

\begin{equation}
\label{ourSigmaEta}
\sigma \approx \frac{Q_3-Q_1}{2 \cdot \Phi^{-1} \left( \frac{0.75 n - 0.125}{n+0.25} \right)  + \exp \left( \frac{n}{-2.882 - 0.231 \cdot n} \right) }
\end{equation}

\bigskip
Observing that:

$$ \lim_{n \to +\infty} \exp \left( \frac{n}{a + b \cdot n} \right) = \exp(1/b) $$

\noindent one concludes that  $\hat \varepsilon(n)$ converges to $\exp(1/(-0.2307863 ...)) = 0.01312794 ...$ . Moreover, as one can verify that: 

$$ \sup_{2 \leq n \leq 50} \left| \eta(n) - 2 \cdot \Phi^{-1} \left( \frac{0.75 n - 0.125}{n+0.25}   \right) \right| \approx 0.580 $$

$$ \inf_{2 \leq n \leq 50} \left| \eta(n) - 2 \cdot \Phi^{-1} \left( \frac{0.75 n - 0.125}{n+0.25}   \right) \right| \approx 0.030 $$

\noindent while:

$$ \sup_{2 \leq n \leq 50} \left| \eta(n) - \left[ 2 \cdot \Phi^{-1} \left( \frac{0.75 n - 0.125}{n+0.25}   \right) + \exp \left( \frac{n}{-2.882 - 0.231 \cdot n} \right) \right] \right| \approx 0.030 $$

$$ \inf_{2 \leq n \leq 50} \left| \eta(n) - \left[ 2 \cdot \Phi^{-1} \left( \frac{0.75 n - 0.125}{n+0.25}   \right) + \exp \left( \frac{n}{-2.882 - 0.231 \cdot n} \right) \right] \right| < 0.0002  $$

\noindent our additive term  (\ref{ourVarepsilon}) allows to gain at least one decimal figure in estimating $\eta(n)$ with respect to original equation (\ref{etaphi}).

\bigskip

\footnotesize The accuracy of $\hat \eta(n)$ can be easily improved. In fact, looking to the diagnostic plot of the linear model which led to relation \ref{ourVarepsilon}, a non-negligible curvature in residuals clearly appears (see \url{https://github.com/MassimoBorelli/sd} for details), inducing not normality and heteroskedasticity. The situation can be amended, particularly if $3 \leq n \leq 50$ with the second order relation: 

$$\hat \varepsilon(n) := \exp  \left( \frac{n}{-9.01647 -0.23238 \cdot (n - 26)  + 0.00074 \cdot (n - 26)^2 } \right)$$

\noindent which is characterized by a residual standard error of 0.035 on 45 degrees of freedom and a multiple R-squared equals to 0.9999, with nearly normal and homoskedastic residuals.
\normalsize

\subsection{Construction of delta(n)}

Let us start noting that the function $ 2 \cdot \Phi^{-1} \left( \frac{n-0.375}{n+0.25} \right) $ proposed by Wan et al. \cite{wan2014estimating} can be already considered a reliable approximation to $\xi(n)$, as

$$ \sup_{2 \leq n \leq 50} \left| \eta(n) - 2 \cdot \Phi^{-1} \left( \frac{n-0.375}{n+0.25}   \right) \right| \approx 0.051 $$

$$ \inf_{2 \leq n \leq 50} \left| \eta(n) - 2 \cdot \Phi^{-1} \left( \frac{n-0.375}{n+0.25}   \right) \right| \approx 0.0003 $$

\noindent Neverthelss, we investigated on the residuals $\delta(n)$ , $2 \leq n \leq 50$  and their approximate derivative $\partial_n$, calculated  by means of the central difference quotient (e.g. cfr. Stoer and Bulirsch \cite{stoer1980introduction}, section 3.5 page 145)

$$\partial_n  = \frac{\delta(n+1) - \delta(n-1)}{2} \; , \; 3 \leq n \leq 49$$

\begin{figure}[ht] 
\label{Arxiv4Plot3}
	\centering
		\includegraphics[scale = 0.4]{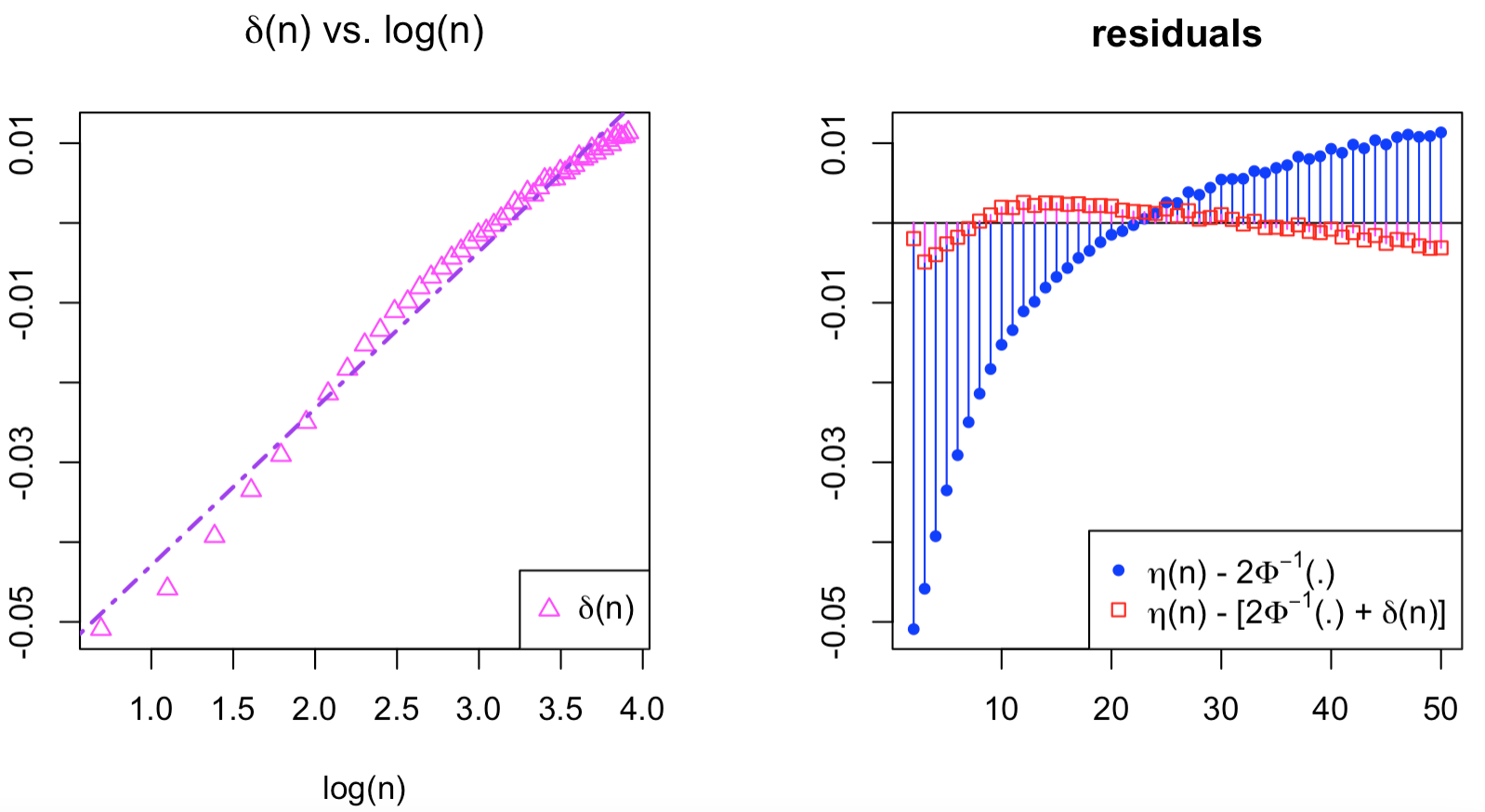} 
\caption{\small On the left, the the dashed purple regression line fitting the linear behaviour of $\hat \delta(n)$ versus $log(n)$ in magenta triangles. On the left, the blue sticks represent the measured $\delta(n)$ residuals, i.e. ${\eta(n) - 2 \cdot \Phi^{-1} \left( \frac{0.75 n - 0.125}{n+0.25} \right) }$; the red squares depict the residuals after correcting with $\hat \delta(n)$, i.e. $\eta(n) - \left( {2 \cdot \Phi^{-1} \left( \frac{0.75 n - 0.125}{n+0.25} \right) } +  0.0197  \cdot \log(n) -0.0626 \right)$. }
\end{figure}

\noindent When plotting on the cartesian plane the sequence $ \frac{1}{\partial_n}$ versus $n$, one can observe an approximate linear behaviour (apart from notable oscillations in the right graph tail). Therefore, we claim that setting:

$$ Y = a + b \cdot \log(n) \equiv \log(A \cdot n^b) $$

\noindent where $a = \log(A)$ one would have:

$$ Y' = 	\frac{d}{dn} Y = 0 + b \cdot \frac{1}{n} =  \frac{b}{n} $$

\noindent that is:

$$ n \propto  \frac{b}{Y'} $$

\noindent i.e. a linear behaviour of $b / Y'$ versus $n$, as actually observed in Figure \ref{Arxiv4Plot3}. Therefore, we set:

\begin{equation}
\label{ourDelta}
\hat \delta(n) := -0.0626 + 0.0197  \cdot \log(n)
\end{equation}

\bigskip

\noindent where $a$ and $b$ were again estimated by the  \texttt{lm} function of \textsf{R} \cite{R},  with a residual standard error of 0.002 and a multiple $R^2 =  0.984$, according to the following summary:

\begin{table}[ht]
\centering
\begin{tabular}{rrrrr}
  \hline
   & Estimate & Std. Error & t value & Pr($>$$|$t$|$) \\ 
  \hline
 $a $ & -0.0626 & 0.0011 & -54.66 & $<$ 0.001 \\ 
  $b$ & 0.0197 & 0.0004 & 53.72 & $<$ 0.001 \\ 
   \hline
\end{tabular}
\end{table}

Consequently, with our proposed approximation (\ref{ourDelta}) for $\delta(n)$, we conclude that in scenario $\mathbf{C_1}$ the function $\xi(n)$ can be approximated by:

\begin{equation}
\label{noiSigmaXi}
\xi(n) \approx \hat \xi(n) := 2 \cdot \Phi^{-1} \left( \frac{n-0.375}{n+0.25} \right) + 0.0197 \log(n) - 0.0626
\end{equation}

\noindent yielding to the improved standard deviation estimating formula:

\begin{equation}
\label{noiSigmaDelta}
\sigma \approx \frac{b-a}{2 \cdot \Phi^{-1} \left( \frac{n-0.375}{n+0.25} \right) + 0.0197 \log(n) - 0.0626 }
\end{equation}

\bigskip 

Lastly, we observe that also in  this case one has approximately an improvement of one order of magnitude in decimal places:

$$ \sup_{2 \leq n \leq 50} \left| \xi(n) - \hat \xi(n) \right| \approx 0.005  $$

$$ \inf_{2 \leq n \leq 50} \left| \xi(n) - \hat \xi(n) \right| < 0.0002  $$

\section{Conclusions}

In conclusion, in this short note we have verified that it is possible to estimate the standard deviation $\sigma$ when knowing the range $\left[ a , b \right]$ and the sample size $n > 1$ according to the following relation:

\begin{equation}
\label{final1}
\sigma = \frac{b-a}{\xi(n)}
\end{equation}

\noindent where:

\[
\xi(n) \approx
\left\{
\begin{array}{ll}
2 \cdot \Phi^{-1} \left( \frac{n-0.375}{n+0.25} \right) + 0.0197 \log(n) - 0.0626  & \mbox{if $ 2 \leq n \leq 50$} \\
2 \cdot \Phi^{-1} \left( \frac{n-0.375}{n+0.25} \right)  & \mbox{if $n > 50$}
\end{array}
\right.
\]

\noindent while, if the quartiles $Q_1$ and $Q_3$ are known, together with the sample size $n > 1$:

\begin{equation}
\label{final2}
\sigma = \frac{Q_3-Q_1}{\eta(n)}
\end{equation}

\noindent where:

\[
\eta(n) \approx
\left\{
\begin{array}{ll}
2 \cdot \Phi^{-1} \left( \frac{0.75 n - 0.125}{n+0.25} \right) + \exp \left( \frac{n}{-2.882 - 0.231 \cdot n} \right) 
  & \mbox{if $ 2 \leq n \leq 50$} \\
2 \cdot \Phi^{-1} \left( \frac{0.75 n - 0.125}{n+0.25} \right)  & \mbox{if $n > 50$}
\end{array}
\right.
\]

\bigskip
\small
\paragraph{Acknowledgments.} M.B. expresses his appreciation to professor Sergio L. Invernizzi of the \textit{Societ\`a dei Naturalisti e Matematici di Modena} for his constructive suggestions, and lifelong teachings.

\bibliographystyle{alpha}
\bibliography{sample}

\end{document}